\documentclass[%
 reprint,
superscriptaddress,
 amsmath,amssymb,
 aps,
]{revtex4-2}

\usepackage{graphicx}
\usepackage{dcolumn}
\usepackage{bm}

\usepackage{upgreek}

\begin{document}

\title{Single-shot on-chip measurement of the modal Stokes parameters}

\author{Amin Hashemi}
\affiliation{CREOL, The College of Optics \& Photonics, University of Central Florida, Orlando, FL 32816, USA}
\affiliation{These authors contributed equally to this work}

\author{Abbas Shiri}
\affiliation{CREOL, The College of Optics \& Photonics, University of Central Florida, Orlando, FL 32816, USA}
\affiliation{These authors contributed equally to this work}

\author{Bahaa E. A. Saleh}
\affiliation{CREOL, The College of Optics \& Photonics, University of Central Florida, Orlando, FL 32816, USA}

\author{Andrea Blanco-Redondo}
\affiliation{CREOL, The College of Optics \& Photonics, University of Central Florida, Orlando, FL 32816, USA}

\author{Ayman F. Abouraddy}
\affiliation{CREOL, The College of Optics \& Photonics, University of Central Florida, Orlando, FL 32816, USA}
\affiliation{raddy@creol.ucf.edu}

\begin{abstract}
Integrated photonics has recently facilitated manipulating multimode partially coherent light in a compact mesh of Mach-Zehnder interferometers. A critical task in this context is on-chip reconstruction of the coherence matrix through a sequence of measurement configurations, thus giving rise to a speed bottleneck stemming from temporal delays in tuning on-chip phase shifts. Here we demonstrate single-shot reconstruction of the coherence matrix for two-mode partially coherent light by parallelizing the on-chip measurements of the modal Stokes parameters, thereby exchanging space for time. Although a larger chip area is required, the reconstruction is limited only by detection speed, thus potentially paving the way to real-time reconstruction of coherence matrices for applications in high-speed communications, cryptography, computation, and sensing.
\end{abstract}


\maketitle

Rapid reconstruction of an unknown, finite-dimensional coherence matrix has recently become essential to fully exploit the `coherence advantage': optical information processing configurations in which partially coherent light outperforms coherent light. Examples include increasing the communications capacity over an $N$-mode channel (e.g., spatial-mode multiplexing in a multimode fiber \cite{Li14AOP}), where partially coherent light provides $\mathcal{O}(N^{2})$ independent communication channels in lieu of $\mathcal{O}(N)$ for coherent light \cite{Nardi22OL}. Another example is coherence-rank optical communications across a rapidly varying, strongly scattering channel \cite{Harling25APLP}, where the rank of the coherence matrix (the number of its non-zero eigenvalues \cite{Harling24PRA,Harling24PRA2}) is a scattering-immune information carrier. Both applications, among other recent examples of optical information processing that also benefit from partially coherent light in computation \cite{Dong24Nature}, cryptography \cite{Liu25LPR}, number factoring \cite{Liu23PRAppl}, and sensing \cite{Tyo06AO}, requiring rapid reconstruction of a coherence matrix $\mathbf{G}$ as a fundamental building block.

To date, two strategies have been utilized to reconstruct $\mathbf{G}$ in on-chip programmable photonics platforms \cite{Capmany20Book,Bogaerts20Nature}. One approach tomographically reconstructs $\mathbf{G}$ by first measuring the modal Stokes parameters (SPs) \cite{James01PRA,Abouraddy02OC,Kagalwala13NP,Abouraddy14OL,Kagalwala15SR}, a generalization of the usual polarization SPs \cite{Brosseau06PO,Wolf07Book} to other degrees-of-freedom of the field (e.g., spatial modes). This strategy sequentially implements a fixed number of measurement settings, realized recently in an on-chip hexagonal mesh of Mach-Zehnder interferometers (MZIs) with two-mode \cite{Hashemi26arxiv} and four-mode \cite{Hashemi26arxiv4Modes} partially coherent light. Alternatively, in `variational processing' pairs of modes traverse a unitary transformation (`unitary' for brevity) whose parameters are scanned to reconstruct $\mathbf{G}$ \cite{roques2024Light}. The speed bottleneck remains the temporal delay encountered when tuning on-chip phase shifters at each step of the measurement sequence, which sets an ultimate limit on the reconstruction speed of $\mathbf{G}$.

Here we demonstrate \textit{single-shot} on-chip reconstruction of the coherence matrix of two-mode partially coherent light using the tomographic reconstruction approach after acquiring the modal SPs. In contrast to recent experiments \cite{Hashemi26arxiv,Hashemi26arxiv4Modes}, we exchange space for time to parallelize the measurements -- rather than sequentially implementing them in a smaller occupied space but with added temporal delays. On-chip detectors acquire the modal weights to estimate the modal SPs and reconstruct $\mathbf{G}$ in a single shot. We confirm that the fidelity of single-shot reconstruction is comparable to that of its sequential counterpart, and we reconstruct an arbitrary $\mathbf{G}$ at a rate limited by only the detector speed. The two-mode single-shot measurement configuration is the building block for larger-dimensional coherence matrices. Our results therefore open a path to real-time, high-speed on-chip reconstruction of the coherence matrix for optical communications and information processing applications. 

Consider an optical field spanned by two modes labeled $|1\rangle$ and $|2\rangle$ in the Dirac notation, which correspond here to the fields in two single-mode on-chip waveguides. The $2\times2$ coherence matrix for this field in terms of the modal SPs is:
\begin{equation}\label{eq:GStokes}
\mathbf{G}=\!\left(\!\begin{array}{cc}G_{11}&G_{12}\\G_{21}&G_{22}\end{array}\!\right)=\frac{1}{2}\sum_{j=0}^{3}s_{j}\hat{\sigma}_{j}=\frac{1}{2}\left(\!\begin{array}{cc}s_{0}+s_{1}&s_{2}-is_{3}\\s_{2}+is_{3}&s_{0}-s_{1}\end{array}\!\right),
\end{equation}
where $G_{jk}=\langle E_{j}E_{k}^{*}\rangle$, $j,k=1,2$, $E_{1}$ and $E_{2}$ are the field amplitudes in the two waveguides, $G_{21}=G_{12}^{*}$, we normalize $\mathbf{G}$ to unity trace $\mathrm{Tr}\{\mathbf{G}\}=G_{11}+G_{22}=1$, $G_{11}$ and $G_{22}$ are the modal weights (the fractions of power in each mode), the real expansion coefficients $\{s_{j}\}_{j=0}^{3}$ are the modal SPs, $\{\hat{\sigma}_{j}\}_{j=1}^{3}$ are the usual Pauli matrices: $\hat{\sigma}_{1}=\footnotesize{\left(\!\!\begin{array}{cc}
1\!\!&\!\!0\\0\!\!&\!\!-1\end{array}\right)}$, $\hat{\sigma}_{2}=\footnotesize{\left(\!\!\begin{array}{cc}0\!\!&\!\!1\\1\!\!&\!\!0\end{array}\!\!\right)}$, and $\hat{\sigma}_{3}=\footnotesize{\left(\!\!\begin{array}{cc}0\!\!&\!\!-i\\i\!\!&\!\!0\end{array}\!\!\right)}$, and $\hat{\sigma}_{0}=\hat{\mathbb{I}}_{2}$ is the $2\times2$ identity matrix. We obtain the modal SPs via $s_{j}=\mathrm{Tr}\{\hat{\sigma}_{j}\mathbf{G}\}$, so that $s_{0}=1$, $s_{1}=G_{11}-G_{22}$, $s_{2}=2\mathrm{Re}\{G_{12}\}$, and $s_{3}=-2\mathrm{Im}\{G_{12}\}$, from which $\mathbf{G}$ is reconstructed (Eq.~\ref{eq:GStokes}). SPs are well known in polarization optics \cite{Wolf07Book} but are equally applicable to any bimodal field \cite{Abouraddy19Optica,Halder21OL}.

\begin{figure}[t!]
\centering
\includegraphics[width=8.5cm]{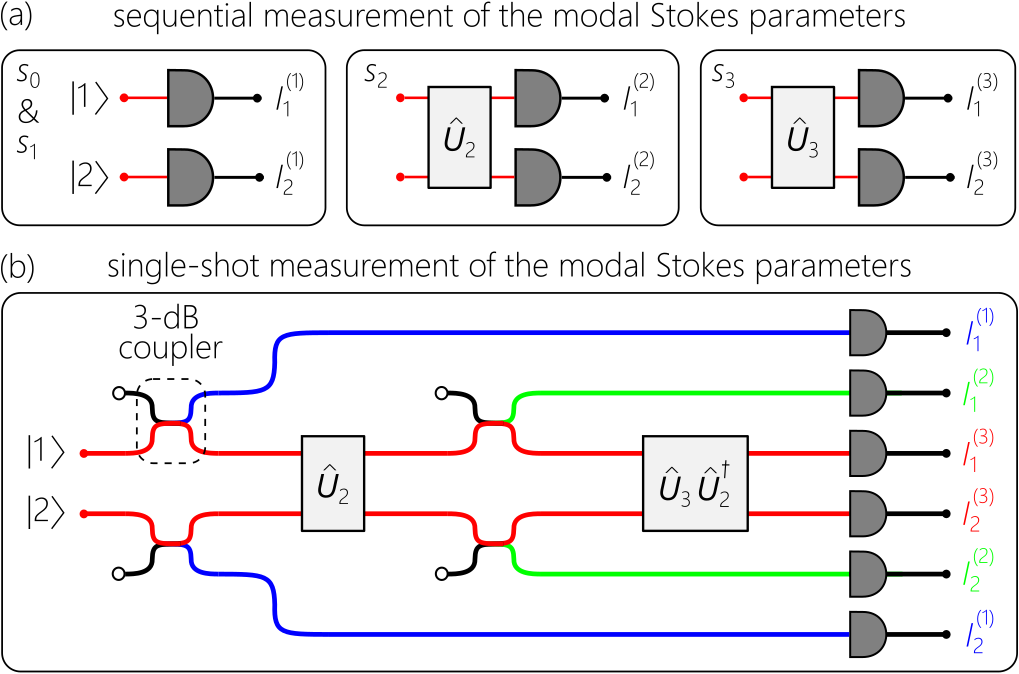} 
\caption{(a) Configurations for the sequential acquisition of the modal SPs: $s_{0}$ and $s_{1}$ using a unitary $\hat{U}_{1}=\hat{\mathbb{I}}_{2}$; $s_{2}$ with $\hat{U}_{2}$; and $s_{3}$ with $\hat{U}_{3}$ \cite{Hashemi26arxiv}. (b) Schematic for the proposed single-shot measurement of the modal SPs.}
\label{fig:BasicStokes}
\end{figure}

To measure $\{s_{j}\}$, we implement a sequence of three settings [Fig.~\ref{fig:BasicStokes}(a)], each involving a unitary $\hat{U}_{j}$ after which the modal weights $I_{1}^{(j)}$ and $I_{2}^{(j)}$ are measured, whereupon $s_{j}=I_{1}^{(j)}-I_{2}^{(j)}$; here, $\hat{U}_{1}=\hat{\mathbb{I}}_{2}$ corresponding to a direct measurement of the modal weights, $\hat{U}_{2}=\tfrac{1}{\sqrt{2}}\footnotesize{\left(\!\!\begin{array}{cc}1\!&\!1\\1\!&\!-1\end{array}\!\!\right)}$, and $\hat{U}_{3}=\tfrac{1}{\sqrt{2}}\footnotesize{\left(\!\!\begin{array}{cc}1\!\!&\!\!-i\\i\!\!&\!\!-1\end{array}\!\!\right)}$. We normalize the SPs with respect to $s_{0}=I_{1}^{(1)}+I_{2}^{(1)}$. Substituting the acquired modal SPs into Eq.~\ref{eq:GStokes} reconstructs $\mathbf{G}$. We refer to this procedure as a `sequential measurement' of the modal SPs. The need to cycle through three unitaries, followed each time with a measurement, indicates that the reconstruction speed is limited by both the on-chip phase shifters and detectors.

\begin{figure}[t!]
\centering
\includegraphics[width=8.5cm]{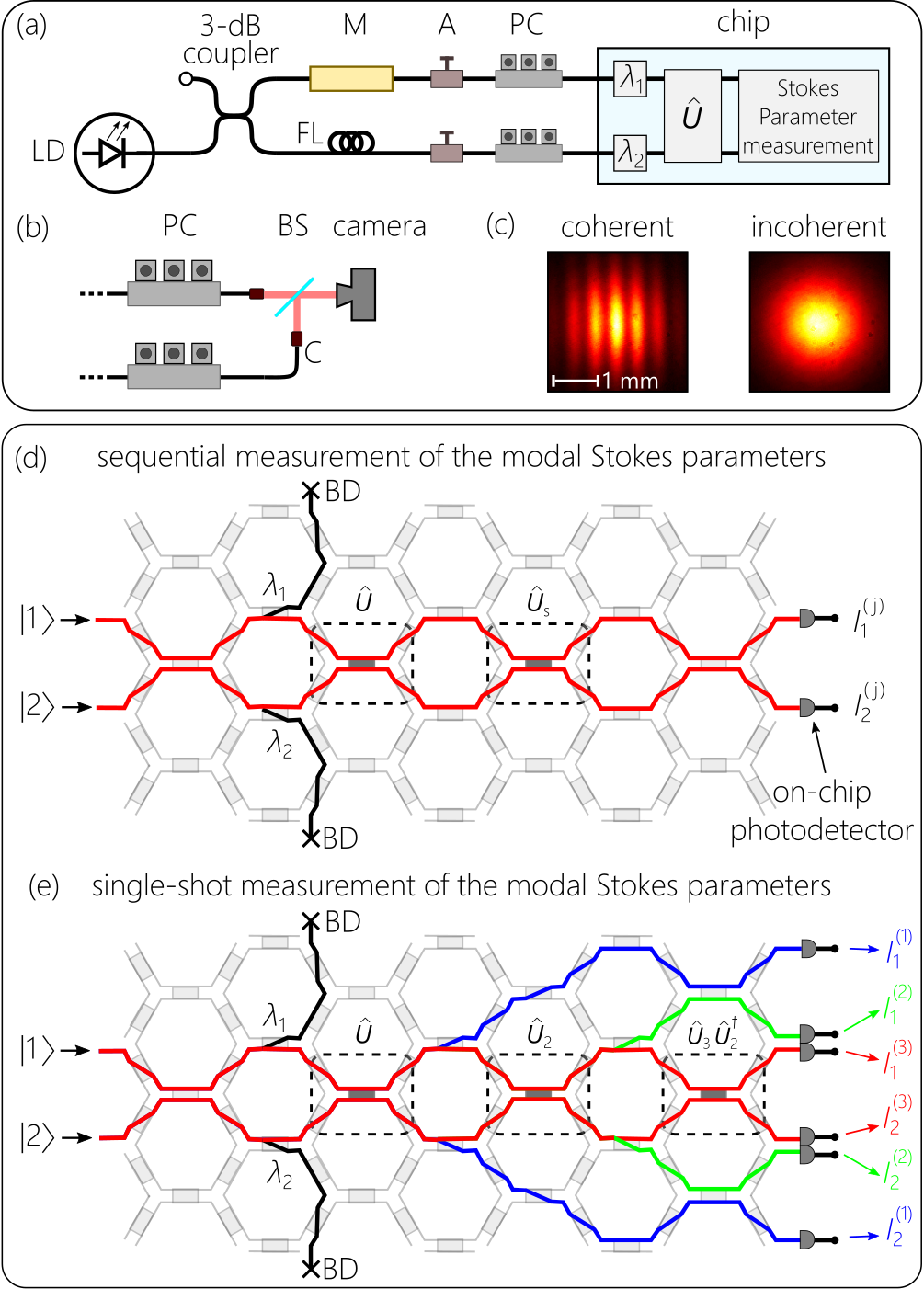} 
\caption{(a) Optical setup. A: Attenuator; PC: polarization controller; FL: fiber loop; M: modulator. Three tasks are performed on chip: adjusting the eigenvalues $\lambda_{1}$ and $\lambda_{2}$, changing the basis of $\mathbf{G}$ via the unitary $\hat{U}$, followed by a measurement of the modal SPs. (b) Setup for recording free-space interference between the two modes. C: Fiber collimator; BS: balanced beam splitter. (c) Intensity profile in absence and presence of the fiber loop in (b), where interference fringes appear or disappear, respectively. (d) Layout of the on-chip MZI mesh for sequential and (g) single-shot measurement of the modal SPs. BD: Beam dump. Each gray rectangle is an MZI and the lines are single-mode on-chip waveguides.}
\label{fig:Setup}
\end{figure}

In the proposed single-shot approach [Fig.~\ref{fig:BasicStokes}(b)], the unitaries are laid out in space concurrently. After each unitary, a portion of the power is diverted by symmetric couplers ($50:50$ split) to detectors. We first record $I_{1}^{(1)}$ and $I_{2}^{(1)}$, from which we obtain $s_{0}=2(I_{1}^{(1)}+I_{2}^{(1)})$ and $s_{1}=2(I_{1}^{(1)}-I_{2}^{(1)})$; after $\hat{U}_{2}$ the detected modal weights $I_{1}^{(2)}$ and $I_{2}^{(2)}$ yield $s_{2}=4(I_{1}^{(2)}-I_{2}^{(2)})$; and after $\hat{U}_{3}'=\hat{U}_{3}\hat{U}_{2}^{\dagger}=\tfrac{1}{\sqrt{2}}e^{-i\tfrac{\pi}{4}}\footnotesize{\left(\!\!\begin{array}{cc}1\!\!&\!\!i\\-1\!\!&\!\!i\end{array}\!\!\right)}$, which undoes $\hat{U}_{2}$ before implementing $\hat{U}_{3}$, the detected modal weights $I_{1}^{(3)}$ and $I_{2}^{(3)}$ yield $s_{3}=4(I_{1}^{(3)}-I_{2}^{(3)})$. The obtained modal SPs $\{s_{j}\}$ are substituted into Eq.~\ref{eq:GStokes} to reconstruct $\mathbf{G}$. In this configuration we exchange space for time, the phase-shifters in the unitaries remain static, and the detectors record the measurements simultaneously. Consequently, the measurement speed is limited solely by the detectors.  This distinction provides the potential for continuous reconstruction of a temporally varying $\mathbf{G}$.

\begin{figure}[t!]
\centering
\includegraphics[width=8.6cm]{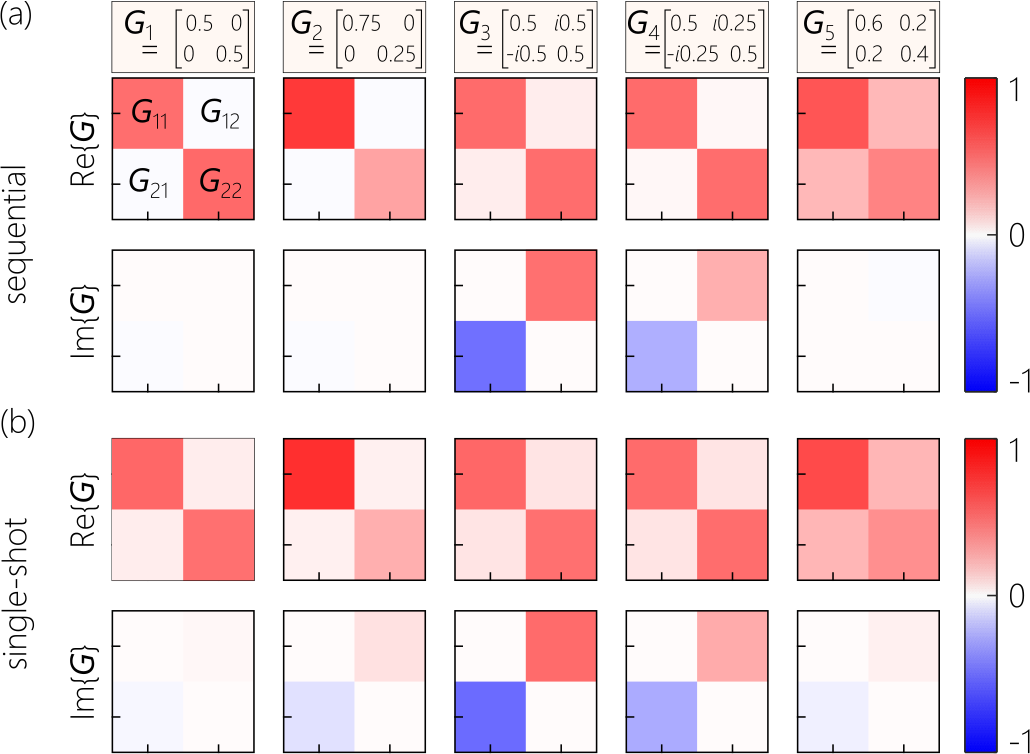} 
\caption{On-chip reconstruction of the coherence matrix $\mathbf{G}$ via measurements of the modal SPs. The columns correspond to the coherence matrices $\mathbf{G}_{1}$ through $\mathbf{G}_{5}$ given in the top row. (a) Reconstructing $\mathbf{G}$ via on-chip (a) sequential and (b) single-shot measurements. We plot $\mathrm{Re}\{\mathbf{G}\}$ and $\mathrm{Im}\{\mathbf{G}\}$ separately. The color map is modified to accentuate even minor differences.}
\label{fig:G}
\end{figure}

We carry out our experiments using the setup in Fig.~\ref{fig:Setup}(a). A laser diode at a wavelength $\approx1.55$~$\mu$m is coupled to a single-mode fiber (SMF), split into two paths using a 3-dB fiber coupler, and a fiber loop of length $L\approx500$~m (exceeding the laser coherence length) is placed in one of the paths. Attenuators and polarization controllers yield equal amplitudes and optimally match the polarization to the on-chip waveguides. To confirm that the SMF fields are mutually incoherent, we out-couple the fields to free space via fiber collimators, superpose the fields at a balanced beam splitter, and record the intensity distribution with a camera (Bobcat-320-GigE-13907, Xenics) [Fig.~\ref{fig:Setup}(b)]. In absence of the fiber loop, the two fields are mutually coherent and interference fringes are observed; while in presence of the fiber loop, the fields are mutually incoherent, and the fringes are eliminated [Fig.~\ref{fig:Setup}(c)]. Because the two modal weights are equal and the modes are mutually incoherent, the prepared coherence matrix is $\mathbf{G}=\tfrac{1}{2}\hat{\mathbb{I}}_{2}$. Once coupled to the chip, we first tune the relative modal weights to $\lambda_{1}$ and $\lambda_{2}$, $\mathbf{G}^{\mathrm{D}}\rightarrow\footnotesize{\left(\!\!\begin{array}{cc}\lambda_{1}\!\!&\!\!0\\0\!\!&\!\!\lambda_{2}\end{array}\!\!\right)}$ with $\lambda_{1}+\lambda_{2}=1$. A unitary $\hat{U}$ then converts this diagonal coherence matrix to any form desired, $\mathbf{G}=\hat{U}\mathbf{G}^{\mathrm{D}}\hat{U}^{\dagger}$, while maintaining its degree of coherence $D=\lambda_{1}-\lambda_{2}$. The degree of coherence is not simply the observable interference visibility when superposing the two modes, but corresponds to the \textit{maximum} visibility observable after equalizing the modal amplitudes \cite{Zernicke,Abouraddy19Optica}; see an experimental demonstration of this concept using on-chip generated two-mode fields in Ref.~\cite{Hashemi26arxiv}. Finally, the coherence matrix is reconstructed on chip via a measurement of the modal SPs -- utilizing both the sequential and single-shot approaches.

The integrated photonic chip (iPronics Smartlight Processor) is a hexagonal mesh of 72~MZIs, each formed of two phase shifts $\varphi_{1}$ and $\varphi_{2}$ for modes $|1\rangle$ and $|2\rangle$, respectively, sandwiched between symmetric couplers, yielding the unitary $\hat{U}_{\mathrm{MZI}}=ie^{i\varphi}\footnotesize{\left(\!\!\begin{array}{cc}\sin\tfrac{\delta}{2}\!\!&\!\!\cos\tfrac{\delta}{2}\\\cos\tfrac{\delta}{2}\!\!&\!\!-\sin\tfrac{\delta}{2}\end{array}\!\!\right)}$, where $\varphi=\tfrac{1}{2}(\varphi_{1}+\varphi_{2})$ and $\delta=\varphi_{1}-\varphi_{2}$\cite{Hashemi26arxiv4Modes,Hashemi26arxiv}. By sandwiching an MZI between relative phase shifters by $\alpha$ and $-\alpha$ between the modes, we obtain a general unitary $\hat{U}=ie^{i\varphi}\footnotesize{\left(\!\!\begin{array}{cc}\sin\tfrac{\delta}{2}\!\!&\!\!e^{-i\alpha}\cos\tfrac{\delta}{2}\\e^{i\alpha}\cos\tfrac{\delta}{2}\!\!&\!\!-\sin\tfrac{\delta}{2}\end{array}\!\!\right)}$ \cite{Hashemi26arxiv,Hashemi26arxiv4Modes}. The chip layouts for the sequential and single-shot schemes are illustrated in Fig.~\ref{fig:Setup}(d,e). In the former, the unitary $\hat{U}_{\mathrm{s}}$ cycles between three selections: $\hat{U}_{1}=\hat{\mathbb{I}}_{2}$ to obtain $s_{1}=I_{1}^{(1)}-I_{2}^{(1)}$, $\hat{U}_{2}$ to obtain $s_{2}=I_{1}^{(2)}-I_{2}^{(2)}$, and $\hat{U}_{3}$ to obtain $s_{3}=I_{1}^{(3)}-I_{2}^{(3)}$. The measurements are acquired at three instances in time following the change in the unitary. In the single-shot realization, unitaries $\hat{U}_{2}$ and $\hat{U}_{3}'=\hat{U}_{3}\hat{U}_{2}^{\dagger}$ are cascaded. Portions of the modes are first diverted to yield $I_{1}^{(1)}$ and $I_{2}^{(1)}$, and after $\hat{U}_{2}$ to yield $I_{1}^{(2)}$ and $I_{2}^{(2)}$. Measurements after $\hat{U}_{3}'$ yield $I_{1}^{(3)}$ and $I_{2}^{(3)}$. The six outputs $I_{1}^{(j)}$ and $I_{2}^{(j)}$, $j=1,2,3$, are recorded simultaneously (single-shot).

We first validate the performance of the single-shot measurement of modal SPs with respect to its sequential counterpart. For this measurement we do not make use of the external modulator, and start from the coherence matrix $\mathbf{G}=\tfrac{1}{2}\hat{\mathbb{I}}_{2}$ coupled to the chip, adjust the eigenvalues $\lambda_{1}$ and $\lambda_{2}$, and then rotate the basis via a unitary $\hat{U}$. We implemented 5~coherence matrices ($\mathbf{G}_{1}$ through $\mathbf{G}_{5}$) and reconstructed them using the sequential and single-shot approaches. We plot in Fig.~\ref{fig:G} the reconstructed coherence matrices via the two approaches, separating the real and imaginary parts, $\mathrm{Re}\{\mathbf{G}\}$ and $\mathrm{Im}\{\mathbf{G}\}$, respectively. We make use of the fidelity $\mathcal{F}=(\mathrm{Tr}\{\sqrt{\mathbf{G}_{\mathrm{m}}}\mathbf{G}_{\mathrm{th}}\sqrt{\mathbf{G}_{\mathrm{m}}}\})^{2}$ as a measure of the reconstruction quality \cite{Jozsa94JMO}, where $\mathbf{G}_{\mathrm{th}}$ and $\mathbf{G}_{\mathrm{m}}$ are the theoretical and reconstructed coherence matrices, respectively. The sequential reconstruction has $\mathcal{F}>99.5\%$ and the single-shot counterpart has $\mathcal{F}\approx99\%$.  

\begin{figure}[t!]
\centering
\includegraphics[width=8.6cm]{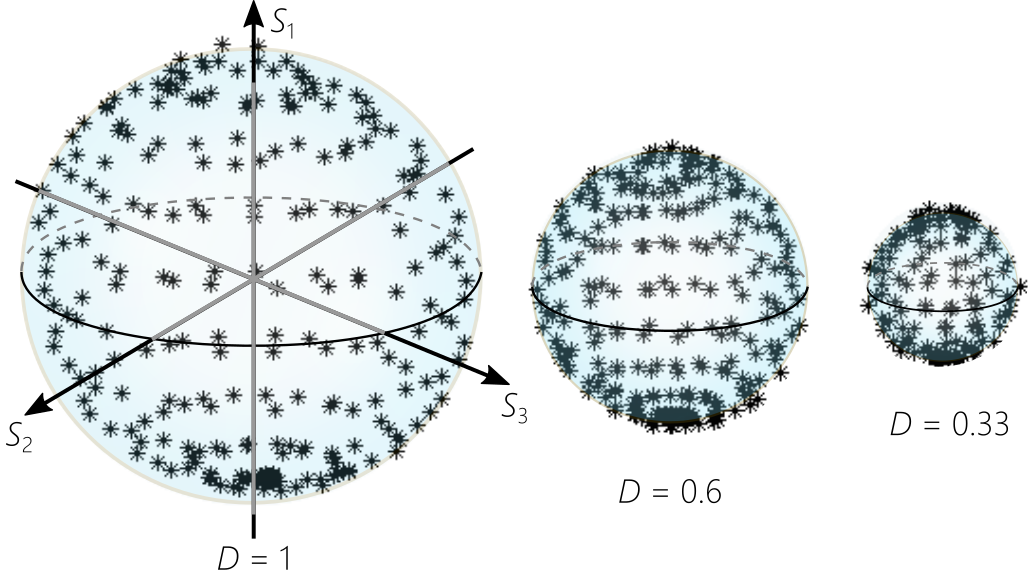}
\caption{Single-shot reconstruction of $\mathbf{G}$. We plot each reconstructed field as a point on the surface of the Poincar{\'e} sphere of radius $D$ and axes $\{s_{1},s_{2},s_{3}\}$, for $D=1$ (coherent fields), $D=0.6$, and $D\approx0.33$.}
\label{fig:PS}
\end{figure}

To ensure that the single-shot approach is capable of reconstructing \textit{any} coherence matrix, we select three values of $D$ ($D=1.,0.6,$ and 0.33) and then sweep the two parameters in the unitary $\hat{U}$ placed in the path of the modes [Fig.~\ref{fig:Setup}(a,d,e)]: the relative phase $\alpha\in[0,2\pi]$ and coupling ratio $\delta\in[0,\tfrac{\pi}{2}]$ introduced between the two modes. This enables spanning the entire surface of the Poincar{\'e} sphere [Fig.~\ref{fig:PS}], thereby confirming that there are no `blind spots' in this detection scheme. The data plotted in Fig.~\ref{fig:PS} indicates that the single-shot approach can reconstruct $\mathbf{G}$ at any point \textit{on} the Poincar{\'e} sphere surface (coherent fields) or \textit{within} its volume (partially coherent fields).

\begin{figure}[t!]
\centering
\includegraphics[width=8.6cm]{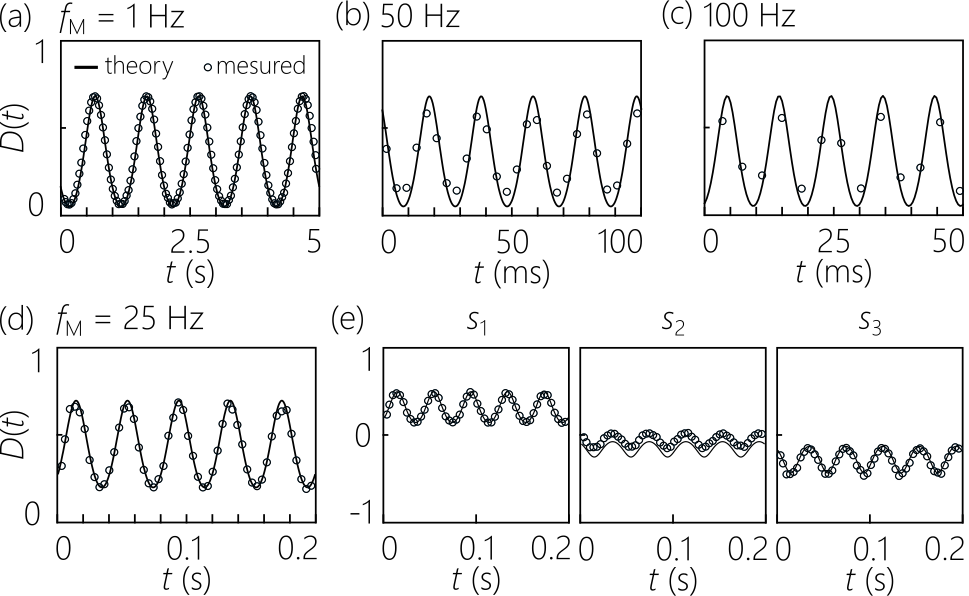}
\caption{Real-time reconstruction of the coherence matrix via single-shot measurement of the modal SPs. (a-c) $D(t)$ obtained from the reconstructed diagonal coherence matrix $\mathbf{G}^{\mathrm{D}}$: (a) $f_{\mathrm{M}}=1$~Hz, (b) 50~Hz, and (c) 100~Hz. (d) $D(t)$ obtained from the non-diagonal reconstructed $\mathbf{G}$ at $f_{\mathrm{M}}=25$~Hz. (e) The acquired individual modal SPs $s_{1}(t)$, $s_{2}(t)$, and $s_{3}(t)$ from which we obtain $D(t)$ as plotted in (d). Data points are spaced 4~ms apart (the on-chip detector speed limit).}
\label{fig:Time}
\end{figure}

Finally, to ascertain the gain in processing speed, we employ an external amplitude modulator (Uniphase) in one input path preceding the chip [M in Fig.~\ref{fig:Setup}(a)]. An electrical signal in the form of a sinusoid of frequency $f_{\mathrm{M}}$ from a waveform generator with an appropriate electrical bias is sent to the modulator to produce the diagonal coherence matrix $\mathbf{G}^{\mathrm{D}}=\tfrac{1}{1+a(t)}\footnotesize{\left(\!\!\begin{array}{cc}a(t)\!\!&\!\!0\\0\!\!&\!\!1\end{array}\!\!\right)}$, where $a(t)$ is the modulated optical power. The eigenvalues are $\lambda_{1}(t)=\tfrac{a(t)}{1+a(t)}$ and $\lambda_{2}(t)=\tfrac{1}{1+a(t)}$, so that the degree of coherence is $D(t)=\tfrac{1-a(t)}{1+a(t)}$. We first set the on-chip unitary to $\hat{U}=\hat{\mathbb{I}}_{2}$, so the coherence matrix to be reconstructed is $\mathbf{G}^{\mathrm{D}}$. In this case, the modal SPs are $s_{1}(t)=\tfrac{1-a(t)}{1+a(t)}$ and $s_{2}(t)=s_{3}(t)=0$, so that $D(t)=\sqrt{s_{1}^{2}(t)+s_{2}^{2}+s_{3}^{2}}=s_{1}(t)$. We carry out single-shot reconstruction of $\mathbf{G}$, from which we extract $D(t)$ [Fig.~\ref{fig:Time}(a-c)]. We vary $f_{\mathrm{M}}$ from 0.1~Hz to 100~Hz, and the on-chip detectors sample the field every 4~ms (the on-chip detector limit). In the sequential approach, the limit on the speed of reconstructing $\mathbf{G}$ is dominated by the speed of on-chip thermal phase-shift change (it is expected that electro-optic phase shifters will be faster). Reconstruction even at $f_{\mathrm{M}}=1$~Hz could not be achieved by the sequential approach. We plot in Fig.~\ref{fig:Time}(a-c) the measured $D(t)$ using the single-shot approach for $f_{\mathrm{M}}=1$, 50, and 100~Hz. This approach continues to reconstruct $\mathbf{G}$ to at least $f_{\mathrm{M}}=25$~Hz. 

Next, we set $\alpha=\tfrac{\pi}{3}$ and $\delta=\tfrac{\pi}{4}$ in $\hat{U}$ to render all the SPs non-zero without changing the degree of coherence $D(t)$. We plot $D(t)$ for $f_{\mathrm{M}}=25$~Hz in Fig.~\ref{fig:Time}(d). The expected SPs are $s_{1}(t)=\tfrac{1}{\sqrt{2}}\tfrac{1-a(t)}{1+a(t)}$, $s_{2}(t)=-s_{1}(t)\cos\tfrac{\pi}{3}$, and $s_{3}(t)=-s_{1}(t)\sin\tfrac{\pi}{3}$, which are in excellent agreement with the measured SPs [Fig.~\ref{fig:Time}(e)]. These results indicate that the single-shot approach allows for real-time reconstruction of $\mathbf{G}$ by first acquiring the SPs.  

We have realized single-shot on-chip measurements of the modal SPs cascaded in space rather than sequenced in time. Other spatial configurations can be pursued that may require fewer MZIs in their construction. The integrated photonic circuit used in our experiments is a general purpose programmable mesh of MZIs, and its topology is not optimized for the particular task at hand. More efficient chip-level layouts can thus further compactify the device size, which will be essential for scaling up the single-shot approach to $N>2$ modes \cite{Hashemi26arxiv4Modes}. 

The single-shot approach alleviates the speed bottleneck arising from on-chip phase shifters, leaving only the detector speed to limit the reconstruction rate. Using high-speed detectors, whether external or on-chip, we can readily exceed a MHz reconstruction rate. Using the reversible conversion between spatial and polarization modes demonstrated in \cite{Abouraddy17OE,Okoro17Optica,Harling22OE,Harling23JO}, we expect integrated photonic systems to enable compact, MHz-rate Stokes polarimetry, which may open new pathways in metrology, biosensing, and molecular control \cite{Constant18JB,Ramella20JO,He21Light}.

In conclusion, we have demonstrated a scheme for single-shot measurements of the modal SPs implemented in an integrated photonic circuit to reconstruct the coherence matrix for two-mode partially coherent light. Rather than sequentially reconfiguring the on-chip circuit serially in time, we instead parallelize the requisite configurations in space to acquire the SPs in a single shot. The on-chip layout is thus static, thereby swapping time for space, or sacrificing the size of the device for higher-speed reconstruction of the coherence matrix -- limited only by the detector speed. These results may help the development of high-speed Stokes polarimetry, in addition to further increasing the speed of on-chip processing of partially coherent light in optical communications, computation, and spectroscopy.\\

Acknowledgments: 
We thank G.~Li and D.~Hudson for loan of equipment and Saikat Saha for conectorizing fibers.\\

Funding: 
U.S. Office of Naval Research (ONR) N00014-20-1-2789.

\bibliography{diffraction}

@ARTICLE{Kagalwala13NP,
  AUTHOR =       {K. H. Kagalwala and G. {Di G}iuseppe and A. F. Abouraddy and B. E. A. Saleh},
  TITLE =        {Bell's measure in classical optical coherence},
  JOURNAL =      {Nat. Photon.},
  YEAR =         {2013},
  volume =       {7},
  pages =        {72-78},
}

@ARTICLE{Abouraddy14OL,
  AUTHOR =       {A. F. Abouraddy and K. H. Kagalwala and B. E. A. Saleh},
  TITLE =        {Two-point optical coherency matrix tomography},
  JOURNAL =      {Opt. Lett.},
  YEAR =         {2014},
  volume =       {39},
  pages =        {2411-2414},
}

@ARTICLE{Kagalwala15SR,
  AUTHOR =       {K. H. Kagalwala and H. E. Kondakci and A. F. Abouraddy and B. E. A. Saleh},
  TITLE =        {Optical coherency matrix tomography},
  JOURNAL =      {Sci. Rep.},
  YEAR =         {2015},
  volume =       {5},
  pages =        {15333},
}

@ARTICLE{Okoro17Optica,
  AUTHOR =       {C. Okoro and H. E. Kondakci and A. F. Abouraddy and K. C. Toussaint},
  TITLE =        {Demonstration of an optical-coherence converter},
  JOURNAL =      {Optica},
  YEAR =         {2017},
  volume =       {4},
  pages =        {1052-1058},
}

@article{James01PRA,
  title = {Measurement of qubits},
  author = {D. F. V. James and P. G. Kwiat and W. J. Munro and A. G. White},
  journal = {Phys. Rev. A},
  volume = {64},
  pages = {052312},
  year = {2001},
}

@BOOK{Wolf07Book,
  AUTHOR =       {E. Wolf},
  TITLE =        {Introduction to the Theory of Coherence and Polarization of Light},
  PUBLISHER =    {Cambridge Univ. Press},
  YEAR =         {2007},
  address =      {Cambridge}
}

@ARTICLE{Nardi22OL,
  AUTHOR =       {A. Nardi and S. Divitt and M. Rossi and F. Tebbenjohanns and A. Militaru and M. Frimmer and L. Novotny},
  TITLE =        {Encoding information in the mutual coherence of spatially separated light beams},
  JOURNAL =      {Opt. Lett.},
  YEAR =         {2022},
  volume =       {47},
  pages =        {4588-4591},
}

@ARTICLE{Abouraddy17OE,
  AUTHOR =       {A. F. Abouraddy},
  TITLE =        {What is the maximum attainable visibility by a partially coherent electromagnetic field in {Y}oung's double-slit interference?},
  JOURNAL =      {Opt. Express},
  YEAR =         {2017},
  volume =       {25},
  pages =        {18320-18331},
}

@ARTICLE{Harling24PRA,
  title = {Locked entropy in partially coherent optical fields},
  author = {M. Harling and V. A. Kelkar and K. C. Toussaint and A. F. Abouraddy},
  journal = {Phys. Rev. A},
  volume = {109},
  pages = {L021501},
  year = {2024},
}

@article{Harling24PRA2,
  title = {Isoentropic partially coherent optical fields that cannot be interconverted unitarily},
  author = {M. Harling and V. A. Kelkar and K. C. Toussaint and A. F. Abouraddy},
  journal = {Phys. Rev. A},
  volume = {110},
  issue = {1},
  pages = {013505},
  numpages = {11},
  year = {2024},
  month = {Jul},
  publisher = {American Physical Society},
}

@ARTICLE{Halder21OL,
  AUTHOR =       {A. Halder and A. Norrman and A. T. Friberg},
  TITLE =        {Poincar{\'e} sphere representation of scalar two-beam interference under spatial unitary transformations},
  JOURNAL =      {Opt. Lett.},
  YEAR =         {2021},
  volume =       {46},
  pages =        {5619-5622},
}

@article{roques2024Light,
  title={Measuring, processing, and generating partially coherent light with self-configuring optics},
  author={C. Roques-Carmes and S. Fan and D. Miller},
  journal={Light Sci. Appl.},
  year={2024},
  volume =       {13},
  pages =        {260},
}

@ARTICLE{Bogaerts20Nature,
  AUTHOR =       {W. Bogaerts and D. P{\'e}rez and J. Capmany and D. A. B. Miller and J. Poon and D. Englund and F. Morichetti and A. Melloni},
  TITLE =        {Programmable photonic circuits},
  JOURNAL =      {Nature},
  YEAR =         {2020},
  volume =       {586},
  pages =        {207–216},
}

@ARTICLE{Li14AOP,
  AUTHOR =       {G. Li and N. Bai and N. Zhao and C. Xia},
  TITLE =        {Space-division multiplexing: the next frontier in optical communication},
  JOURNAL =      {Adv. Opt. Photon.},
  YEAR =         {2014},
  volume =       {6},
  pages =        {413-487},
}

@ARTICLE{Brosseau06PO,
  AUTHOR =       {C. Brosseau and A. Dogariu},
  TITLE =        {Symmetry properties and polarization descriptors for an arbitrary electromagnetic wavefield},
  JOURNAL =      {Prog. Opt.},
  YEAR =         {2006},
  volume =       {49},
  pages =        {315-380},
}

@ARTICLE{Zernicke,
  AUTHOR =       {F. Zernicke},
  TITLE =        {The concept of the degree of coherence and its application to optical problems},
  JOURNAL =      {Physica},
  YEAR =         {1938},
  volume =       {5},
  pages =        {785-795},
}

@article{Harling22OE,
  title={Reversible inter-degree-of-freedom optical-coherence conversion via entropy swapping},
  author={M. Harling and V. Kelkar and C. Okoro and M. Diouf and A. F. Abouraddy and K. C. Toussaint},
  journal={Opt. Express},
  volume={30},
  pages={29584-29597},
  year={2022},
}

@ARTICLE{Harling23JO,
  AUTHOR =       {M. Harling and V. Kelkar and A. F. Abouraddy and K. C. Toussaint},
  TITLE =        {Reversible coherence conversion across optical degrees-of-freedom: a tutorial},
  JOURNAL =      {J. Opt.},
  YEAR =         {2023},
  volume =       {25},
  pages =        {053502},
}

@ARTICLE{Harling25APLP,
  AUTHOR =       {M. Harling and C. Stevenson and K. C. Toussaint and A. F. Abouraddy},
  TITLE =        {Optical communications through highly scattering channels using the coherence-rank},
  JOURNAL =      {APL Photonics},
  YEAR =         {2025},
  volume =       {10},
  pages =        {076116},
}

@ARTICLE{Abouraddy19Optica,
  AUTHOR =       {A. F. Abouraddy and A. Dogariu and B. E. A. Saleh},
  TITLE =        {Comment on `Polarization coherence theorem'},
  JOURNAL =      {Optica},
  YEAR =         {2019},
  volume =       {6},
  pages =        {829-830},
}

@ARTICLE{Dong24Nature,
  AUTHOR =       {B. Dong and F. Br{\"u}ckerhoff-Pl{\"u}ckelmann and L. Meyer and J. Dijkstra and I. Bente and D. Wendland and A. Varri and S. Aggarwal and N. Farmakidis and M. Wang and G. Yang and J. S. Lee and Y. He and E. Gooskens and D.-L. Kwong and P. Bienstman and W. H. P. Pernice and H. Bhaskaran},
  TITLE =        {Partial coherence enhances parallelized photonic computing},
  JOURNAL =      {Nature},
  YEAR =         {2024},
  volume =       {632},
  pages =        {55-62},
}

@ARTICLE{Liu25LPR,
  AUTHOR =       {X. Liu and X. Li and S. A. Ponomarenko and F. Wang and X. Peng and Y. Cai and C. Liang},
  TITLE =        {Unlocking secure optical multiplexing with spatially incoherent light},
  JOURNAL =      {Laser Photon. Rev.},
  YEAR =         {2025},
  volume =       {19},
  pages =        {2401534},
}

@ARTICLE{Jozsa94JMO,
  AUTHOR =       {R. Jozsa},
  TITLE =        {Fidelity for mixed quantum states},
  JOURNAL =      {J. Mod. Opt.},
  YEAR =         {1994},
  volume =       {41},
  pages =        {2315-2323},
}

@Book{Capmany20Book,
  AUTHOR =       {J. Capmany and D. P{\'e}rez},
  TITLE =        {Programmable Integrated Photonics},
  Publisher =    {Oxford Univ. Press},
  YEAR =         {2020},
  address =      {Oxford},
}

@ARTICLE{Hashemi26arxiv,
  AUTHOR =       {A. Hashemi and A. Shiri and B. E. A. Saleh and A. Blanco-Redondo and A. F. Abouraddy},
  TITLE =        {Programmable on-chip synthesis and reconstruction of partially coherent two-mode optical fields},
  JOURNAL =      {arXiv:2601.09802},
  YEAR =         {2026},
  volume =       {},
  pages =        {},
}

@ARTICLE{Abouraddy02OC,
  AUTHOR =       {A. F. Abouraddy and A. V. Sergienko and B. E. A. Saleh and M. C. Teich},
  TITLE =        {Quantum entanglement and the two-photon Stokes parameters},
  JOURNAL =      {Opt. Commun.},
  YEAR =         {2002},
  volume =       {210},
  pages =        {93-98},
}

@ARTICLE{Hashemi26arxiv4Modes,
  AUTHOR =       {A. Hashemi and A. Shiri and B. E. A. Saleh and A. Blanco-Redondo and A. F. Abouraddy},
  TITLE =        {On-chip control of the coherence matrix of four-mode partially coherent light: rank, entropy, and modal Stokes parameters},
  JOURNAL =      {arXiv:2601.18797},
  YEAR =         {2026},
  volume =       {},
  pages =        {},
}

@ARTICLE{Constant18JB,
  AUTHOR =       {C. Constant and A. Bergano and K. Sugaya and A. Dogariu},
  TITLE =        {Guiding cellular activity with polarized light},
  JOURNAL =      {J. Biophoton.},
  YEAR =         {2018},
  volume =       {11},
  pages =        {e201600326},
}

@ARTICLE{He21Light,
  AUTHOR =       {C. He and H. He and J. Chang and B. Chen and H. Ma and M. J. Booth},
  TITLE =        {Polarisation optics for biomedical and clinical applications: a review},
  JOURNAL =      {Light Sci. Appl.},
  YEAR =         {2021},
  volume =       {10},
  pages =        {194},
}

@ARTICLE{Ramella20JO,
  AUTHOR =       {J. C. Ramella-Roman and I. Saytashev and M. Piccini},
  TITLE =        {A review of polarization-based imaging technologies for clinical and preclinical applications},
  JOURNAL =      {J. Opt.},
  YEAR =         {2020},
  volume =       {22},
  pages =        {123001},
}

@ARTICLE{Tyo06AO,
  AUTHOR =       {J. S. Tyo and D. L. Goldstein and D. B. Chenault and J. A. Shaw},
  TITLE =        {Review of passive imaging polarimetry for remote sensing applications},
  JOURNAL =      {Appl. Opt.},
  YEAR =         {2006},
  volume =       {45},
  pages =        {5453-5469},
}

@ARTICLE{Liu23PRAppl,
  AUTHOR =       {X. Liu and C. Liang and Y. Cai and S. A. Ponomarenko},
  TITLE =        {Axial correlation revivals and number factorization with structured random waves},
  JOURNAL =      {Phys. Rev. Appl.},
  YEAR =         {2023},
  volume =       {20},
  pages =        {L021004},
}

\end{document}